\documentclass[journal]{IEEEtran}

\usepackage{xcolor,soul,framed} 

\colorlet{shadecolor}{yellow}
\usepackage[pdftex]{graphicx}
\graphicspath{{../pdf/}{../jpeg/}}
\DeclareGraphicsExtensions{.pdf,.jpeg,.png}

\usepackage[cmex10]{amsmath}
\usepackage{array}
\usepackage{mdwmath}
\usepackage{mdwtab}
\usepackage{eqparbox}
\usepackage{caption}
\usepackage{url}
\usepackage{cite}
\usepackage{algorithmic}
\usepackage{graphicx}
\usepackage{textcomp}
\usepackage{xcolor}

\usepackage{xcolor,soul,framed} 
\usepackage{adjustbox}
\usepackage{tabularx}
\colorlet{shadecolor}{yellow}

\usepackage{graphicx}
\usepackage{amssymb}
\usepackage[cmex10]{amsmath}
\usepackage{array}
\usepackage{mdwmath}
\usepackage{mdwtab}
\usepackage{eqparbox}
\usepackage{url}
\usepackage{multirow}
\usepackage{subfig}
\usepackage{graphicx}
\usepackage{multirow}
\usepackage{amsfonts}
\usepackage{cite}

\usepackage{graphicx}

\begin{document}

    \title{Data-Driven Chance Constrained AC-OPF \\using Hybrid Sparse Gaussian Processes}
  \author{Mile Mitrovic, 
          Aleksandr Lukashevich,
          Petr Vorobev,
          Vladimir Terzija,
          Yury Maximov,
          Deepjyoti Deka
  \thanks{Mile Mitrovic, Aleksandr Lukashevich, Petr Vorobev and Vladimir Terzija are with Skolkovo Institute of Science and Technology, 143025, Moscow, Russia, $\{$mile.mitrovic, aleksandr.lukashevich, p.vorobev, v.terzija$\}$@skoltech.ru; Deepjoyti Deka and Yury Maximov are with Los Alamos National Laboratory, 87545, Los Alamos NM, USA,  $\{$deepjoyti, yury$\}$@lanl.gov.}
  }

\maketitle

\begin{abstract}
The alternating current (AC) chance-constrained optimal power flow (CC-OPF) problem addresses the economic efficiency of electricity generation and delivery under generation uncertainty. The latter is intrinsic to modern power grids because of the high amount of renewables. Despite its academic success, the AC CC-OPF problem is highly nonlinear and computationally demanding, which limits its practical impact. For improving the AC-OPF problem complexity/accuracy trade-off, the paper proposes a fast data-driven setup that uses the sparse and hybrid Gaussian processes (GP) framework to model the power flow equations with input uncertainty. We advocate the efficiency of the proposed approach by a numerical study over multiple IEEE test cases showing up to two times faster and more accurate solutions compared to the state-of-the-art methods. 
\end{abstract}


\section{Introduction}
The optimal power flow (OPF) problem is a fundamental tool for the secure and economic operation of power system, used in system security assessments and in electricity markets. Nowadays, the increase of renewable energy generation is seen as a primary approach to reduce greenhouse emissions \cite{hockstad2018inventory}. 
Despite the benefits of renewable energy sources (RES), the uncertainty associated with RES generation such as wind and solar leads to significant control challenges for the Transmission System Operators (TSOs)~\cite{liu2012challenges}, \cite{koutsoyiannis2016unavoidable}. Optimal optimization and feedback control algorithms should act on second and sub-second levels to prevent the grid from impending energy blackout due to RES uncertainty/fluctuations. Various methods address the accuracy and scalability of the AC OPF problem under uncertainty, including seminal probabilistic optimal power flow \cite{morshed2018probabilistic, shargh2016probabilistic}, and chance-constrained approach \cite{bienstock2014chance, du2021chance, viafora2020chance, lukashevich2021importance,owen2019importance,  zhang2011chance, schmidli2016stochastic, roald2017chance, lubin2019chance, roald2013analytical}. The chance-constrained setup considered in this paper ensures that the security constraints are satisfied with a high probability and allows TSOs to balance operating costs and system security. Unfortunately, Chance-Constrained (CC) AC-OPF with non-linear Alternating Current-Power flow (AC-PF) equations is known to be computationally demanding and often does not have an analytical solution \cite{cousins2014cubic}, \cite{khachiyan1993complexity}. A simpler linear version, chance-constraint direct current (DC) approximation (DC CC-OPF) \cite{bienstock2014chance, du2021chance,roald2017chance} has been widely studied in literature and has a few scalable solutions \cite{viafora2020chance,lukashevich2021power}, but suffers from conservative solutions. Furthermore, it requires knowledge of the system parameters such as line impedances that may have errors and uncertainities. 

The paper proposes to improve the AC CC-OPF scalability by using a data-driven method that builds upon the framework of Gaussian Process (GP) Regression to replace the power flow equations using prior historical data. It is worth mentioning that GP based power-flow models have been implemented in deterministic OPF formulations~\cite{pareek2020gaussian, zamzam2020learning, pan2020deepopf,kekatos}, and for CC-OPF in \cite{mile}. Compared to vanilla Gaussian process regression in \cite{mile}, however, our approach uses (1) a \textbf{hybrid GP} that combines linear DC-PF with data-driven estimation of the residuals between DC-PF and AC-PF, and (2) uses \textbf{sparse} Gaussian  regression using few selective data for the estimation of the hybrid GP. Both these steps significantly reduce our approach computational complexity without compromising the solution's accuracy, making it amenable for real-time/online deployment. We demonstrate a practical advantage of our approach against state-of-the-art algorithms on multiple IEEE test-cases. 

This paper is organized as follows. Section \ref{sec:2} describes the problem addressed and the uncertainty modeling. Section \ref{sec:3} outlines the Gaussian Process Regression and their sparse version. The hybrid approach to solve the AC CC-OPF problem is given in Section~\ref{sec:4}.  Section \ref{sec:5} presents results of numerical evaluation over various IEEE test cases. Section \ref{sec:6} presents a critical discussion, possible extensions and a summary of the paper.

\section{Background and Uncertainty Modeling} \label{sec:2} 

\subsection{Grid Uncertainty}
Let $\mathcal{Q}(\mathcal{B},\mathcal{E}$) be a  transmission grid with a set of buses $\mathcal{B}$, $b = |{\cal B}|$, and (directed) lines $\mathcal{E}$, $e = |{\cal E}|$. 
We decompose each power injection at a bus into a controllable generation (referred as $g$), uncertain part $r$, and load $l$:
\begin{gather}\label{eq:1}
    p = p_g - p_l + p_r, \qquad q = q_g - q_l + q_r,
\end{gather}
where $p$ and $q$ are active and reactive power injections resp. 
The classical AC-OPF problem assumes that deterministic power injections are known; however, load and renewable generation fluctuate in practice. Let $\omega$ be an active power fluctuation so that $p(\omega) = p + \omega$, $q(\omega) = q + \gamma \omega$ where $\gamma$ is a constant power factor, $\gamma > 0$.


\subsection{Generation and Voltage Control}
Following uncertainty realizations $\omega$, controllable generators adapt their generation to maintain the total power balance through the automatic generation control (AGC) \cite{roald2017chance}. The AGC assumes that the total power mismatch $\Omega = \sum_{i\in \mathcal{G}} \omega_i$ is divided among the generators according to the participation factors $\alpha$, $\alpha > 0$, as:
\begin{align}
    p_{g,i}(\omega) = p_{g,i} + \alpha_i \Omega,\quad \sum_{i\in \mathcal{G}} \alpha_i = 1,  \; \alpha_i \ge 0, \;  \forall i \in \mathcal{G}. \label{eq:agc}
    \end{align}


\subsubsection{Power Flows}
Assume $p_{ij}$ and $q_{ij}$ denote the active and reactive power flows from bus $i$ to bus $j$ along line $(i, j) \in \mathcal{E}$. In the AC power flow formulation, each transmission line's active and reactive power flows depend non-linearly on the voltage magnitudes $v$ and voltage angles $\theta$. The AC power flow equations are given by:   
\begin{align*}
    p_{ij}(\omega) & = v_i(\omega) v_j(\omega)\left[G_{ij}cos\left[\theta_{ij}(\omega)\right] + B_{ij}sin\left[\theta_{ij}(\omega)\right]\right] \\
    q_{ij}(\omega) & = v_i(\omega) v_j(\omega)\left[G_{ij}sin\left[\theta_{ij}(\omega)\right] - B_{ij}cos\left[\theta_{ij}(\omega)\right]\right]
\end{align*}
where $\theta_{ij} = \theta_i - \theta_j$, $G_{ij}$ and $B_{ij}$ are the entries of the real and imaginary parts of the network admittance. The system is balanced when the power flows leaving each bus are equal to the sum of power injection at that bus. 
The apparent power flow on the line $ij$ is given by $s_{ij} = \sqrt{(p_{ij})^2 + (q_{ij})^2}$.



\subsection{Chance Constraints AC OPF}
The full setup of the AC chance-constrained OPF problem is as follows \cite{roald2017chance}:

\vspace{-.8em}
\begin{subequations}\label{eq:9}
\begin{align}
 &\min_{p_g, q_q, v, \theta} \sum_{i\in \mathcal{G}} \mathbb{E}[c_i(p_{g,i}(w))] \\
 &\text{s.t.~} F(\theta(w), v(w), p(w), q(w)) = 0, ~~~~~~~ \forall w\in W \label{eq:9b} \\
 &~~~~~ \mathbb{P}(p_{g_i}(\omega)  \leq p^{max}_{g_i}) \geq 1 - \epsilon_p, ~~~~~~~~~~ \forall i \in \mathcal{G} \label{eq:9c}\\
 &~~~~~ \mathbb{P}(p_{g_i}(\omega)  \geq p^{min}_{g_i}) \geq 1 - \epsilon_p, ~~~~~~~~~~ \forall i \in \mathcal{G}\\
 &~~~~~ \mathbb{P}(q_{g_i}(\omega)  \leq q^{max}_{g_i}) \geq 1 - \epsilon_q, ~~~~~~~~~~ \forall i \in \mathcal{G}\\
 &~~~~~ \mathbb{P}(q_{g_i}(\omega)  \geq q^{min}_{g_i}) \geq 1 - \epsilon_q, ~~~~~~~~~~ \forall i \in \mathcal{G}\\
 &~~~~~ \mathbb{P}(v_{j}(\omega)  \leq v^{max}_{j}) \geq 1 - \epsilon_v, ~~~~~~~~~~~ \forall i \in \mathcal{B}\\
 &~~~~~ \mathbb{P}(v_{j}(\omega)  \geq v^{min}_{j}) \geq 1 - \epsilon_v, ~~~~~~~~~~~ \forall i \in \mathcal{B}\\
 &~~~~~ \mathbb{P}(s_{ij}(\omega)  \leq s^{max}_{ij}) \geq 1 - \epsilon_s, ~~~~~~~~~~ \forall ij \in \mathcal{E} \label{eq:9i}
\end{align}
\end{subequations}
where expectations $\mathbb{E}$ and probabilities $\mathbb{P}$ are taken over an unknown distribution of uncertainty $\omega$, $c_i(\cdot)$ is a cost function of the $i$-th generator. The objective function minimizes the expected total cost of active power generations, with cost coefficients $c$. Eq. (2b) represents AC power balance equations that govern relation between power flows, injections and voltage phasors for all possible fluctuations $\omega$ and buses $i \in \mathcal{B}$. The power outputs of conventional generators, voltage magnitudes at buses, and apparent power flow in the lines are constrained using the different chance constraints (2c)–(2i) that must be satisfied with the prescribed probabilities $\epsilon_p$, $\epsilon_q$, $\epsilon_v$, and $\epsilon_s$. In this way, the chance constraints define an $\epsilon$-reliability set within which an operator is safe to execute additional control actions.




In this paper, we consider separate chance constraints, which limit the violation probability for individual constraints and are easy to enforce \cite{lubin2019chance, lukashevich2021power}. The chance constraints \eqref{eq:9} are intractable for non-linear AC-PF; however, Gaussian Processes allow to design a more tractable approximation to them. The latter can easily be handled with modern optimization solvers.

\section{Model Description} \label{sec:3}

\subsection{Notation}
For a matrix $X$, $X_{ij}$ denotes element $ij$ of a matrix $X$, and $[x]_i$, $[x]_{:,j}$ its $i$-th row and $j$-th column, respectively. With $[x]_{*,j}$, we denote $*$ element of $j$-th column vector. Further, we will use $\cdot_*$ subscript to denote unseen test data. For the given vector $x$, $diag(x)$ is used to refer to a diagonal matrix with entries of vector $x$, while $diag(X)$ represent the vector of diagonal elements of matrix $X$. The probability density of $x$ is denotes as $p(x)$, $p(x|y)$ representing the probability of $x$ for given $y$. $\bigtriangledown(\cdot)_{|x_*}$ and $\bigtriangledown^2(\cdot)_{|x_*}$  are the first and second derivative of the function $(\cdot)$ evaluated at~$x_*$.

\subsection{Problem Formulation}
In the classical AC-OPF modeling with Gaussian processes \cite{mile,pareek2020gaussian}, the Gaussian process regression model infers the full AC-OPF function in Eq.~\eqref{eq:9} from previously collected input and output measurement samples as
\begin{equation*}
    y_i = g(x_i) + \omega_i, ~~~~~ i=1,...,N
\end{equation*}
where $N$ is the total number of samples, $\omega \sim \mathcal{N}(0, \Sigma_{\omega})$ is the Gaussian  noise with covariance $\Sigma_{\omega}$ and $g$ is the GP function. The input $x$ is the vector of controllable active power generators $p_g$, and uncertain active power injections $p_l$ and $p_{r}$. Note that the reactive power injections $q_l$ and $q_r$ are not considered as inputs to the GPR model as they are given by a fixed power factor ratio from the active power injections, increasing redundancy. The output vector $y$ corresponds to all dependent output variables, such as voltages at non-generator buses, reactive power at generator buses, and apparent power flow in the lines. Thus, $x = [p_g^T, p_l^T, p_r^T]$ and $y=[v^T, q_g^T, s^T]$. 


\begin{figure}[t] 
    \centering
    \subfloat[Classical Approach: GP learned on historical data;]{%
        \includegraphics[width=0.24\textwidth]{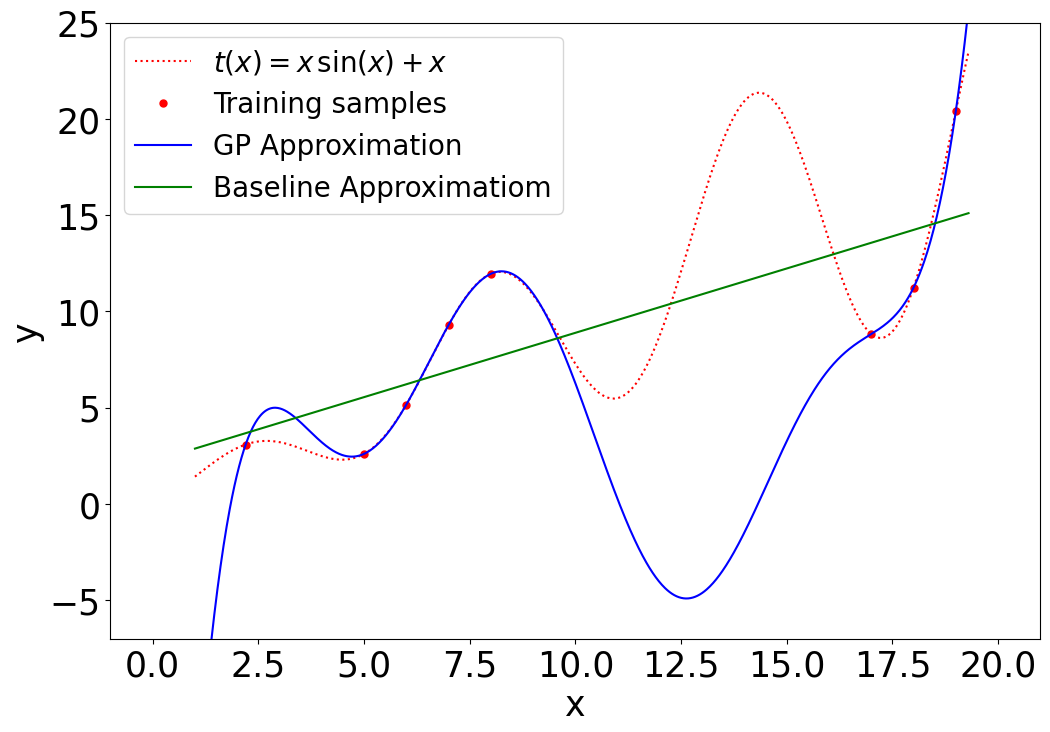}%
        \label{fig:1a}%
        }%
    \hfill%
    \subfloat[Hybrid approach: GP learned on  residuals between linear model and data.]{%
        \includegraphics[width=0.24\textwidth]{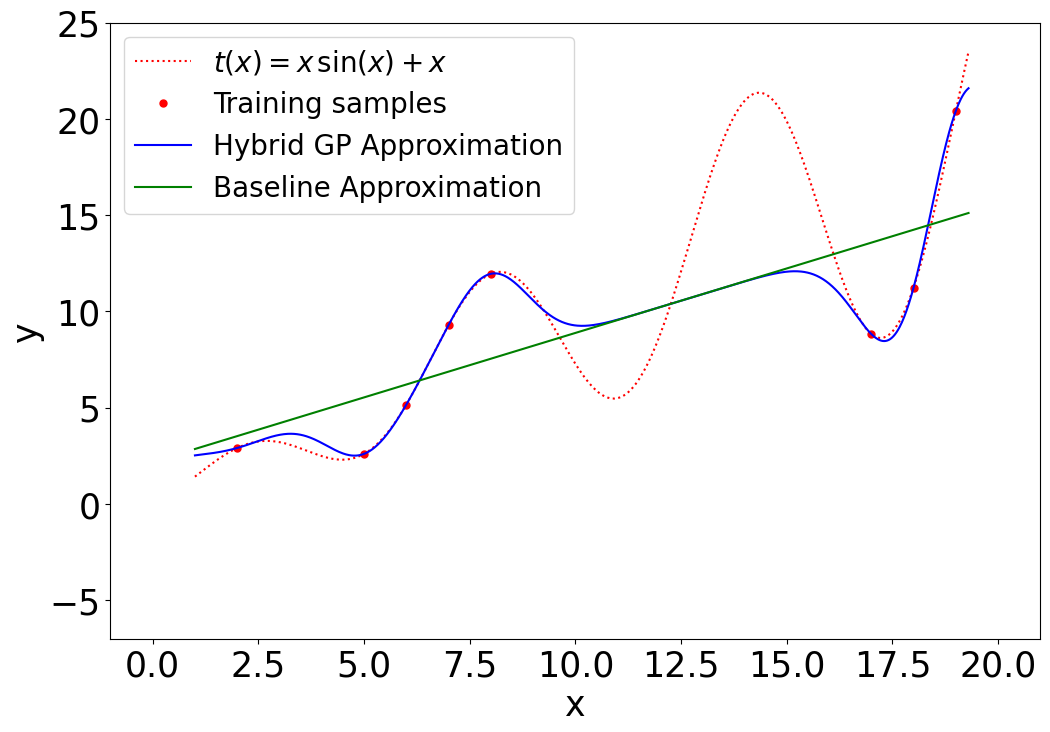}%
        \label{fig:1b}%
        }%
    \caption{Non-linear function $x \sin(x) + x$ (in red) is approximated with GP (blue; left) based on several observations (red dots). The hybrid GP approximation (blue; right) uses a linear proxy (green) to improve the approximation. }
    \label{fig:1}%
\end{figure}

The disadvantage of this approach is its high complexity that requires a lot of samples for an accurate approximation over a linearized fit of the non-linear function. Fig.~\ref{fig:1}(a) illustrates this phenomenon. Our proposed \textit{hybrid} GPR for AC-PF avoids this by combining a linear DC approximated physical model, and an additive GP-part that is learned on the residuals between the non-linear AC and the linear DC model:
\begin{equation}\label{eq:12}
y_i - z(x_i) = g(x_i) + \omega_i, \qquad i=1,...,N,
\end{equation}
where $z$ is a known or estimated linear part, and $g$ is an additive GP-term that describes unknown non-linearities of the system. As shown for a candidate non-linear function in Fig.~\ref{fig:1}(b), this approach has better fit. IN the next section, we describe how the hybrid GP is estimated using available historical data. 

\subsection{Gaussian Process Regression}
Gaussian process regression (GPR) \cite{rasmussen2003gaussian} is a non-parametric Bayesian approach used for non-linear regression problems. We employ GPR, as a supervised machine learning model, to infer a vector-valued residual function $g(x_i)$ in \eqref{eq:12} using a dataset of $N$ samples generated from previously collected measurements of residual outputs $r_i$ and inputs $x_i$ as:
\begin{equation}\label{eq:13}
    r_i = y_i - z(x_i) = g(x_i) + \omega_i, ~~~~~ i=1,...,N
\end{equation}
where the measurement noise on data points $\omega_i$ corresponds to the process noise in \eqref{eq:12}. Historical dataset of inputs $x$ and residual outputs $r$ for learning GPR model is given by:
\begin{equation*}
    S = \{(x_i, r_i)_{i=1}^{N}, x_i\in \mathbb{R}^{n_x}, r_i\in \mathbb{R}^{n_y} \}
\end{equation*}
where $n_x$ and $n_y$ are total number of input and output variables. 

Each output dimension $r_{a}$, $a=1,...,n_y$  is learned independently, given the input data $X = [x_i^T]_{1\leq i\leq N} \in \mathbb{R}^{N \times n_x}$. Thus, specifying a GP-prior on $g$ in each output dimension $a \in \{1, ..., n_y \}$ with defined a prior mean function $m_a(\cdot)$, chosen kernel $k_a(\cdot, \cdot)$ and conditioning on the data S results in a normally (Gaussian) posterior distribution along dimension $a$ at a test point $x_*$, defined with mean and variance:
\begin{subequations}\label{eq:15}
\begin{align}
    \mu_a(x_*) = k^T_*(K_a + \sigma^2_{n,a}I)^{-1}[r]_{:,a} \label{eq:15a}~~~\\
    \sigma^2_a(x_*) = K_{**} - k^T_*(K_a + \sigma^2_{n,a}I)^{-1}k_* \label{eq:15b}
\end{align}
\end{subequations}
where $\sigma^2_{n,a}$ is a variance of \textit{process} noise corresponding to \textit{measurement} noise in \eqref{eq:13}; $I$ is an identity matrix; $[K_a]_{ij} = k_a(x_i, x_j)$ is the Gram matrix of the sampled training data points; $k_* = k_a(x_i, x_*)$ is the vector of covariance between unseen point and training samples; $K_{**} = k_a(x_*,x_*)$ is scalar; and $m_a(X) = [m_a(x_1), m_a(x_2), ..., m_a(x_N) ]$. According to \cite{rasmussen2003gaussian}, a good prior leads to good data interpolation in GPR. We consider a commonly used zero prior mean $m(X) = 0$. Below we use the squared exponential (SE) kernel for the covariance:
\begin{equation*}
 k_a(x_i,x_j) = \sigma^2_{f,a}\exp\left(-\frac{1}{2}(x_i-x_j)^T \Lambda^{-1}_a (x_i-x_j)\right)
\end{equation*}
in which $\sigma^2_{f,a}$ is the signal variance and $\Lambda_a = diag[l^2_{a,1}, l^2_{a,2},..., l^2_{a,{n_x}}]$ is a positive diagonal length scale matrix. The predictive distribution of the multivariate output dimension is given by:
\begin{align*}
    \mu(x_*) & = [\mu_1(x_*), \mu_2(x_*), ..., \mu_{n_y}(x_*)]^T ~~~\\
    \Sigma(x_*) & = diag([\sigma^2_1(x_*), \sigma^2_2(x_*), ..., \sigma^2_{n_y}(x_*)])
\end{align*}
Cholesky decomposition is applied to minimize computational efforts in inverting $(K_a + \sigma^2_{n,a}I)$ during prediction.

GPR training comes down to optimization of the hyperparameters $\theta =[\{\Lambda_a\}, \sigma^2_{f,a}, \sigma^2_{n,a}]$, where $\sigma^2_{n,a}$ is a process noise and $\{\Lambda_a\}$ is the set of length scale hyperparameters in the symmetric matrix $\Lambda_a$. The Maximum Likelihood Estimate with the log marginal likelihood is utilized to optimize hyperparameters as follows \cite{williams2006gaussian}:
\begin{equation}\label{eq:18}
 \begin{split}
 \log~p([r]_{:,a}|X, \Theta) = \frac{1}{2}[r]_{:,a}^{T}(K_a + \sigma^2_{n,a}I)[r]_{:,a} \\ - \frac{1}{2}\log|K_a + \sigma^2_{n,a}I| - \frac{N}{2}\log 2 \pi
 \end{split}
\end{equation}

The negative log likelihood is minimized to find optimal hyperparameters values using SLSQP solver \cite{kraft1988software}. 

\subsection{Uncertainty Propagation}
In the GPR model, for the given deterministic inputs, the predicted outputs have a normal (Gaussian) distribution described with mean and variance \eqref{eq:15}. Since we consider a stochastic problem, the inputs $x$ are also random. Evaluating the outputs of a GPR from an input distribution is generally intractable, and the resulting outputs' distribution is not normal.  Although some approximations of the resulting distribution exist~\cite{koller2018learning}, most of them are computationally demanding and very conservative. Thus, we consider computationally cheap and a practical approximation where the uncertain inputs are modelled via Gaussian distribution:
\begin{equation}\label{eq:19}
    x_* \sim \mathcal{N}(\mu_{x_*}, \Sigma_{x_*}) = \mathcal{N} \left(\begin{bmatrix}
 p_g\\
 p_d
 \end{bmatrix}, 
 \begin{bmatrix}
 \Sigma_g & \Sigma_{gd}\\\
 \Sigma_{gd}^T & \Sigma_d
 \end{bmatrix}\right)\\
\end{equation}
where $p_g$, $p_d = [p^T_l, p^T_r]^T$ are vectors of optimized control variables and forecasted operating points; $\Sigma_g \in \mathbb{R}^{n_g \mathsf{x} n_g}$, $\Sigma_d \in \mathbb{R}^{n_d \mathsf{x} n_d}$ and $\Sigma_{gd} \in \mathbb{R}^{n_g \mathsf{x} n_d}$ are input covariance sub-matrices. 

The covariance $\Sigma_d$ models variances of fluctuations of loads and renewable energy based generators and correlations between them. We assume $\Sigma_d$ to be diagonal, so $(\Sigma_d)_{ii} = \sigma_i^2, (\Sigma_d)_{ij} = 0\; \forall i\neq j$. 

The sub-matrix $\Sigma_g$ refers to uncertainties in generators which are caused due to the fluctuations in loads and RES via Automated Generation Control (AGC) (see Eq.~\ref{eq:agc}). Following the AGC, the covariance in the generator's response is given as
\[
(\Sigma_g)_{ij} = \alpha_i \alpha_j tr(\Sigma_d),\qquad  i, j \in {\cal G}
\]
Similarly, the correlation between the generation and the uncertain inputs $d$ are given by the sub-matrix $\Sigma_{gd}$:
\[(\Sigma_{gd})_{ij} = \alpha_i \sigma_j^2, \quad i \in {\cal G}, j \in {\cal B}\setminus {\cal G}, \]
where $r_t$ is given by Eq.~\eqref{eq:13}.

To determine $\mu_a(x_*)$ and $\sigma^2_a(x_*)$ from a Gaussian input $x_* \sim \mathcal{N}(\mu_{x_*}, \Sigma_{x_*})$, various approximation methods of the posterior of the GPR model have been proposed in the literature \cite{girard2003gaussian, deisenroth2010efficient}. This work considers a first-order Taylor approximation (TA1) for fitting nonlinearities. Applying the first order Taylor approximation (TA1) in Eq.~\eqref{eq:15}, the mean value and variance of output distribution are:
\begin{subequations}\label{eq:20}
\begin{align}
    \mu_{TA1}(x_*) = \mu_a(\mu_{x_*}) \label{eq:20a} ~~~~~~~~~~~~~~~~~~~~\\
    \sigma^2_{TA1}(x_*) = \sigma^2_a(\mu_{x_*}) + \bigtriangledown\mu(\mu_{x_*})\Sigma_{x_*}(\bigtriangledown\mu(\mu_{x_*}))^T
\end{align}
\end{subequations}
The mean prediction on a random input $x_*$ does not provide any correction over the zero-order. At the same time, the variance takes into account the gradient of the posterior mean \eqref{eq:15a} and the covariance $\Sigma_{x_*}$ of the input of the GPR.

\subsection{Sparse Approximation}
Large data sets make the GPR model intractable since the inversion of the $N\times N$ covariance matrix $K_a$ requires $O(N^3)$ time. Sparse Gaussian Processes reduce complexity by selecting the inducing inputs and the kernel hyperparameters by maximizing a lower bound of the marginal log-likelihood~\cite{titsias2009variational}. 
The idea behind sparse Gaussian Processes is to use a small set of points, often referred to as inducing points, $X_m = [x_i^T]_{1\leq i\leq M} \in \mathbb{R}^{M \times n_x}, ~ M \leq N$, to better approximate marginal likelihood. For such a subsample, the approximate posterior GPR mean and covariance function depends on a mean vector $\mu_m$ and a covariance matrix $A_m$ as:
\begin{align*}
    \widehat{\mu}_a(x_*) & = k^T_{m*}(K_{mm,a} + \sigma^2_{m,a}I)^{-1}\mu_{m,a} \\
    \widehat{\sigma}^2_a(x_*) & = K_{**} - k^T_{m*}(K_{mm,a} + \sigma^2_{m,a}I)^{-1}k_{m*} \\
    &\qquad + k^T_{m*}K_{mm,a}A_mK_{mm,a}k_{m*}
\end{align*}
where 
\begin{gather*}
 \mu_{m,a} = \sigma^{-2}_{m,a} K_{mm,a} \Gamma K_{mn,a}[r]_{:,a}, \quad A_m = K_{mm,a} \Gamma K_{mm,a}\\
 \Gamma = (K_{mm,a} + \sigma^{-2}_{m,a} K_{mn,a} K_{nm,a})^{-1}
\end{gather*}
where $K_{mm}$ is $M\times M$ covariance matrix of $\{X_i\}_{i=1}^m$; $K_{mn}$ is $M\times N$ covariance matrix between $\{X_i\}_{i=1}^M$ and the training set $X$; $k_{m*}$ is the cross-covariance vector.  

The sparse approximation reduce the time complexity of the GPR prediction to $O(NM^2)$  from $N^3$ preserving information from original sample span.

\section{Hybrid CC-OPF based on Gaussian Process Regression} \label{sec:4}
\subsection{Setup}
The chance-constrained optimization problem given by Eq.~\eqref{eq:9} is computationally hard, and solving a deterministic problem with over uncertainty realizations instead of the stochastic one is a common way to decrease its complexity. Although the letter helps to get rid of uncertainty, the non-linearity of the power flow equations remains an issue. 

In contrast to the classical scenario optimization approach mentioned above, modeling power flow equations with uncertainty through Gaussian processes lead to a more tractable solution. 
In our hybrid approach, we substitute the power flow balance equation with a combination of a linear approximation  and an additive GPR model learned on the residuals:
\begin{subequations}\label{eq:23}
\begin{align}
    \mu_{y} = f(\mu_{x_*}) + \mu(x_*) ~~~~~~~~~~\\
    \sigma^2_{y} = diag(\Sigma_f(\Sigma_{x_*})) + diag(\Sigma(x_*))
\end{align}
\end{subequations}
where $f(\mu_{x_*}) = [v_{dc}, A\mu^T_{x_*} + b]$; $v_{dc} = 1$ p.u. is constant DC output voltage, $A$ and $b$ are the parameters; $diag(\Sigma_f(\Sigma_{x_*}))$ and $diag(\Sigma(x_*))$ are column vectors of diagonal matrix elements where 
$\Sigma_f = (0,  0; 0 A \Sigma_{x_*} A^\top)$.
To improve tractability, one can reformulate chance constraints for both output and input decision variables. According to~\cite{hewing2019cautious}, one can express individual output probabilities in terms of the mean $\mu_y$ and variance $\sigma^2_y$ as: 
\begin{subequations}\label{eq:24a}
\begin{align}
 \begin{cases}
    \mathbb{P}(\mu_{y_z} + \sqrt{\sigma^2_{y_z}} \leq y_z^{max}) \leq 1-\epsilon_{y_z}\\
      \mathbb{P}(\mu_{y_z} - \sqrt{\sigma^2_{y_z}} \geq y_z^{min}) \leq 1-\epsilon_{y_z}.
 \end{cases}
\end{align}
\end{subequations}
equivalently $y_z^{\min} + \tau_{y_z} \sqrt{\sigma^2_{y_z}} \leq \mu_{y_z} \leq y_z^{\max} - \tau_{y_z} \sqrt{\sigma^2_{y_z}}$, 
where $\tau_{y_z}$ is the quantile function $\Phi^{-1}(1-\epsilon_{y_z})$ of the standard normal distribution and $z=1,...,n_y$.

Similarly, one can reformulate constraints on the input variables as in \cite{bienstock2014chance}:
\begin{subequations}\label{eq:25a}
\begin{align}
 \begin{cases}
    \mathbb{P}(p_{g_k} + \alpha_k\, tr(\Sigma_d)\leq p_{g_k}^{max}) \leq 1-\epsilon_{p_{g_k}}\\
      \mathbb{P}(p_{g_k} - \alpha_k\, tr(\Sigma_d) \geq p_{g_k}^{min}) \leq 1-\epsilon_{p_{g_k}}
 \end{cases}
\end{align}
\end{subequations}
equivalently, as in \eqref{eq:24a}:
\begin{gather}
     p_{g_k}^{min} + \tau_{p_{g_k}} \alpha_k\, tr(\Sigma_d) \leq p_{g_k} \leq p_{g_k}^{max} - \tau_{p_{g_k}} \alpha_k\, tr(\Sigma_d)
\end{gather}
where $\tau_{p_{g_k}} = \Phi^{-1}(1-\epsilon_{p_{g_k}})$ and $k=1,...,n_g$.

The cost function depends quadratically on the participation factors $\alpha$ and generator active powers $p_g$: 
\begin{equation}\label{eq:26} 
\begin{aligned}
 \sum_{k\in \mathcal{G}}\mathbb{E}_{\omega}&[c_k(p_{g,k}(w))] \\
 &= \sum_{k\in \mathcal{G}} \{ c_{2k}(p_{g_k}^{2} + tr(\Sigma_d)\alpha_{k}^{2}) + c_{1k}p_{g_k} + c_{0k}\} \\
 \end{aligned}
\end{equation}
where $\{c_{2k}, c_{1k}, c_{0k}\}_{k=1}^{n_u} \geq 0 $ are scalar cost coefficients.

Applying the GP-balance equation \eqref{eq:23}, an analytical reformulation of chance-constraints with uncertainty margins \eqref{eq:24a}~\eqref{eq:25a}, the hybrid GP CC-OPF problem is:
\begin{subequations}\label{eq:27}
\begin{align}
 &\min_{p_g, \alpha, \mu_y, \sigma_y} \sum_{k\in  \mathcal{G}} \{ c_{2k}(p_{g_k}^{2} + tr(\Sigma_d)\alpha_{k}^{2}) + c_{1k}p_{g_k} + c_{0k}\} \\
 &\text{s.t.~~} \sum_{k \in \mathcal{G}} \alpha_k = 1, ~ \alpha_k \geq 0 \\
 &~~~~~ \sum_{k\in \mathcal{G}} p_{g_k} = \sum_{i\in n_l} p_{l_i} - \sum_{j\in n_r} p_{r_j} \label{eq:27c} \\
 &~~~~~~  \mu_{y} = f(\mu_{x_*}) + \mu(x_*)  \label{eq:27d}\\ 
 &~~~~~~ \sigma^2_{y} = diag(\Sigma_f(\Sigma_{x_*})) + diag(\Sigma(x_*))  \label{eq:27e}\\
 &~~~~~~ y_z^{min} + \lambda_{y_z} \leq \mu_{y_z} \leq y_z^{max} - \lambda_{y_z}\\
 &~~~~~~ u_k^{min} + \lambda_{p_{g_k}} \leq p_{g_k} \leq u_k^{max} - \lambda_{p_{g_k}} 
\end{align}
\end{subequations}
where $\lambda_{y_z} = \tau_{y_z} \sqrt{\sigma^2_{y_z}}$ for $z\in n_y$ and $\lambda_{p_{g_k}} = \tau_{p_{g_k}} \alpha_k\, \sqrt{tr(\Sigma_d)}$ are the uncertainty margins; $\sqrt{tr(\Sigma_d)}$ represents the standard deviation of the total active power imbalance $\Omega$. The equations \eqref{eq:27c}--\eqref{eq:27e} describe the hybrid GP-balance equations by replacing a AC power flow equations with a data-driven regression vector-function. These equations represent the balance between generation and consumption in the power systems by optimizing the decision variables of the controllable generators. However, the proposed OPF optimization method \eqref{eq:27} is a non-convex optimization problem. Since the SE kernel is twice differentiable, the problem can be solved using a nonlinear optimization method. Accordingly, we used a primal-dual interior-point algorithm with filter line-search to solve \eqref{eq:27}. This optimization problem is solved within the CasADi~\cite{andersson2019casadi} nonlinear optimization framework using the available IPOPT~\cite{wachter2006implementation} solver.

\subsection{Evaluation Metrics}
To evaluate the performance of the GPR regression model, the root mean square error (RMSE) metric is used for each output variable $a$. If the predicted output vector $\hat{y}$ (GP-mean values) and ground truth $y$ (AC-PF output values), the RMSE is calculated as follows:
\begin{equation}\label{eq:28}
 RMSE_a = \sqrt{\frac{1}{N_*}\sum_{i=1}^{N_*} ([y]_{i,a} - [\hat{y}]_{i,a})^2}
\end{equation} 
where $N_*$ is number of test samples. The average RMSE score of all outputs $RMSE = n^{-1}_y\sum_{a=1}^{n_y} RMSE_a$ is the accuracy measure we use in our experiments.

\section{Results} \label{sec:5}
We used Intel Core i7-5500U CPU @ 2.40GHz with 8GB RAM for numerical computations and \texttt{pandapower} \cite{thurner2018pandapower} package to validate the results. The source code for experiments is accessible online\footnote{\url{https://github.com/mile888/hybrid_gp}}.

\subsection{Case Study}
Performance and scalability of the proposed hybrid GP CC-OPF method are evaluated over 9-bus and 39-bus IEEE test cases. We model the forecast errors as multivariate Gaussian random variables with zero mean and standard deviation corresponding to 15\% of (forecasted) loads ($\sigma_l = 0.15\, p_l$) and 30\% for the renewable generations ($\sigma_r = 0.3\, p_r$). Following \cite{abbaspourtorbati2015swiss}, we set the acceptable violation probabilities of generator limits to $\epsilon_{p_g}=0.1\%$ for controllable generators, and $\epsilon_y = 2.5\%$ for all other constraints. 

\subsection{Model Performance}
The hybrid and full GPR models are trained on residuals between datasets collected from full AC-OPF and DC-OPF approximation. The synthetic data of the IEEE 9-bus system consists of 8 input features and 15 outputs, while for IEEE 39, the model has 37 inputs and 74 outputs. We generated 75 and 200 random training samples for IEEE 9 and 39-bus systems and considered a different number of inducing points of 10, 30, and 50 for both full and hybrid GPR models.  The iterative interior-point algorithm to solve both hybrid and full GPR CC-OPF and find a feasible solution with convergence tolerance of $\epsilon_{tol} = 10^{-5}$. 

Models with the TA1 approximation method taking into account the full training dataset and a different number of inducing points 
are compared using RMSE as an accuracy score and CPU time to estimate computational complexity.

The results in Table \ref{table:table1} indicate that the hybrid GPR model with TA1 outperforms both RMSE and CPU time with respect to full TA1~\cite{mile}. 
Similarly, sparse approximation allows to significantly reduce the time complexity for both hybrid and full approaches. 
%
Notice, that RMSE computed between the DC approximation and the set of samples is RMSE = $6.65e^{-2}$ p.u. (IEEE 9) and $3.04e^{-2}$ p.u. (IEEE 39). In other words, applying the GPR model with a few samples leads to at least 6 times more accurate approximation than the DC baseline. 


\begin{table}[ht]
\centering
 \begin{tabular}{l|l|l|l|l}
 {\bf Test Case} &
 \multicolumn{2}{c}{\bf IEEE 39} & \multicolumn{2}{c}{\bf IEEE 9}\\
 \hline
 Parameters &  RMSE & Time &  
 RMSE & Time\\
 & $10^{-2}$ p.u. & sec. &  $10^{-2}$ p.u. & sec.\\
 \hline
 hybrid TA1 &  \textbf{0.53} & 10.82&  \textbf{0.30} & 0.24\\
 \hline
 hybrid-sparse 50 TA1 &  1.12 & 2.93 &  0.56 & 0.10  \\
 \hline
 hybrid-sparse 30 TA1 & 1.22 & 1.23 & 0.79 & 0.06 \\
 \hline
 hybrid-sparse 10 TA1 &  1.41 & \textbf{0.51} &  1.22 & \textbf{0.02}\\
 \hline
 full TA1 &  1.37 & 19.38 & 0.65 & 0.35 \\
 \hline
 full-sparse 50 TA1 & 2.69 & 4.12 & 0.97 & 0.17\\
 \hline
 full-sparse 30 TA1 & 2.80 & 1.42 & 1.67 & 0.13 \\
 \hline
 full-sparse 10 TA1 & 3.25 & \textbf{0.51}  & 2.43 & 0.03 
 \end{tabular}
 \caption{GP CC-OPF approximation method results for IEEE 9 and IEEE 39 bus systems} 
 \label{table:table1}
\end{table}



 

\subsection{Comparison of GP CC-OPF to Sample-Based Methods}

In this section, we compare solution cost, computational time, and reliability of the proposed hybrid and full GPR CC-OPF approaches with Scenario Approximation (SA) CC-OPF \cite{mezghani2020stochastic}. In all cases, we use the first-order Taylor approximation (TA1) as a function linearization to simplify the computations. The empirical constraint violation is computed using 1000 Monte-Carlo samples that follow (Gaussian) uncertainty distribution. 

We compare our approach to two baselines. (A) we solve a deterministic AC-OPF problem for each uncertainty realization and take a corresponding quantile in the output distribution. Although computationally efficient, this approach is not guaranteed to provide a (probabilistically) feasible solution; (B) once solving a deterministic AC-OPF, one gets we get optimal participation factors $\alpha^*$ and power distribution. Later on, we find a solution to the AC power flow balance equation for each uncertainty realization and run the Automatic Generation Control with fixed participation factors. 

Notice that (A) considers generalized nonlinear feedback, while (B) does not optimize the participation
factor $\alpha$ and uses the same participation factors for each sample. Thus they give the upper and the lower bound on the feasible solution value. Numerical evaluation of the bounds in given in Table~\ref{table:table2}.

\begin{table}[ht]
\centering
\begin{adjustbox}{width=.45\textwidth}
 \begin{tabular}{l|l|l|l|l|l|l}
 System &
 \multicolumn{3}{c}{IEEE 39} & \multicolumn{3}{c}{IEEE 9} \\
 \hline
 Parameters & Cost, & Failure & Time& Cost, & Failure & Time\\
  & $10^6$ & Prob., \% & sec.& $10^3$ & Prob., \% & sec.\\
 \hline
 A (full recourse) & 7.87 & n/a & 0.93& 4.06 & - & 0.86\\
 \hline
 B (base case)& 7.53 & 35.36 & 0.93 & 3.47 & 9.76 & 0.86\\
 \hline
 Full GP CC-OPF & 7.75 & 2.36 & 19.38 & 4.04 & 0.44 & 0.35\\
 \hline
 Hybr. GP CC-OPF & 7.75 & 2.48 & 9.58 & 4.02 & 0.8 & 0.12\\
 \hline
 CC-OPF (20 scenarios) &n/a& n/a& n/a & 3.48 & 7.52 & 4.4\\
 \hline
 CC-OPF (50 scenarios) & 7.64 & 14.20 & 96.4 & 3.84 & 4.92 & 12.1\\
 \hline
 CC-OPF (100 scenarios) & 7.70 & 8.24 & 184.70 & 3.99 & 1.24 & 22.80 \\
 \hline
 CC-OPF (200 scenarios)& 7.81 & 0.16 & 505.00 & n/a & n/a &n/a
 \end{tabular}
 \end{adjustbox}
 \caption{Cost function values, CPU time and probability of a failure for various setups.} 
 \label{table:table2}
\end{table}



\subsection{Comparison of Uncertainty Margins}
\begin{figure}[ht]
 \centering
 \includegraphics[width=4cm]{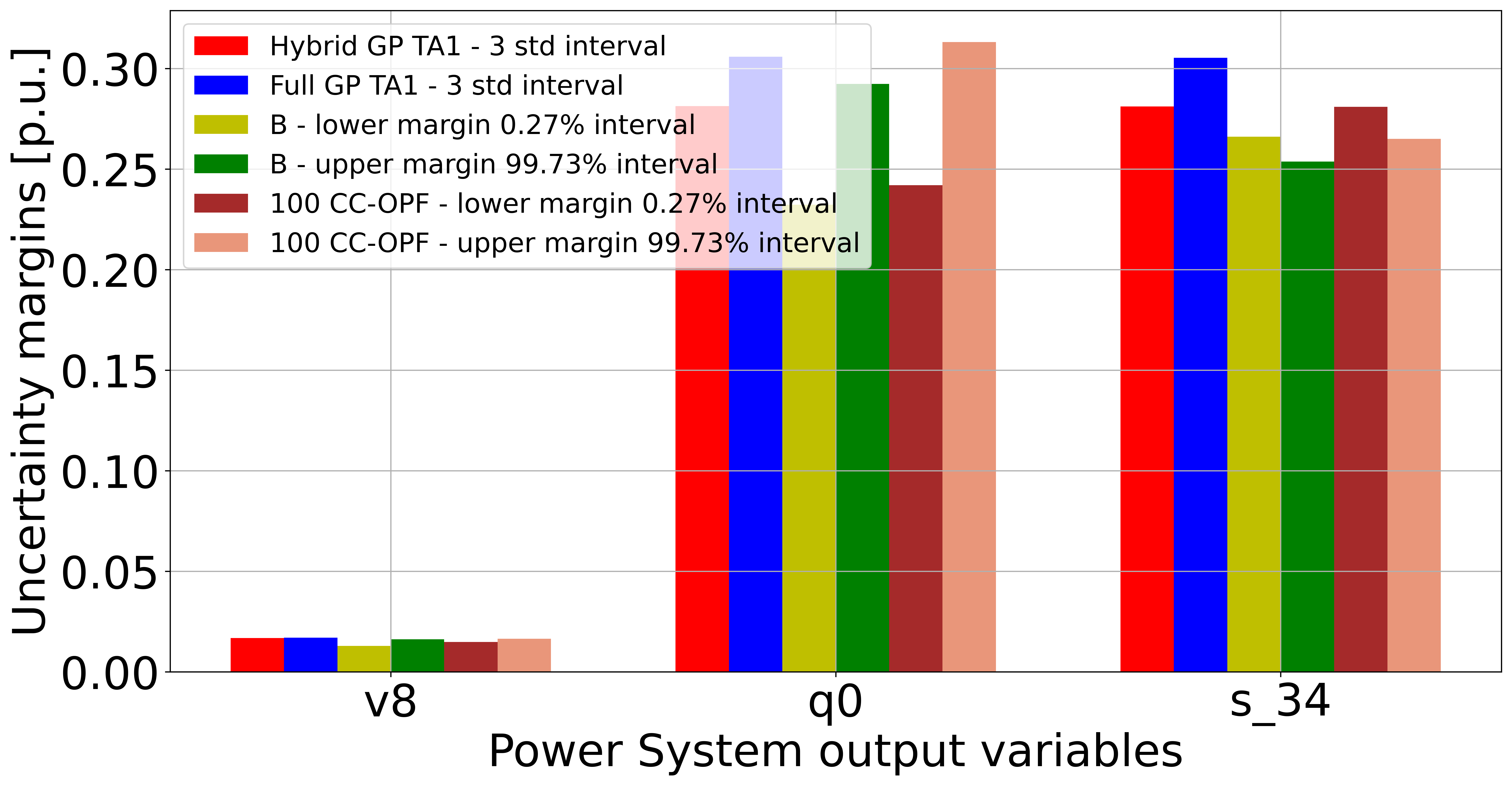} \hspace{0.5cm}
 \includegraphics[width=4cm]{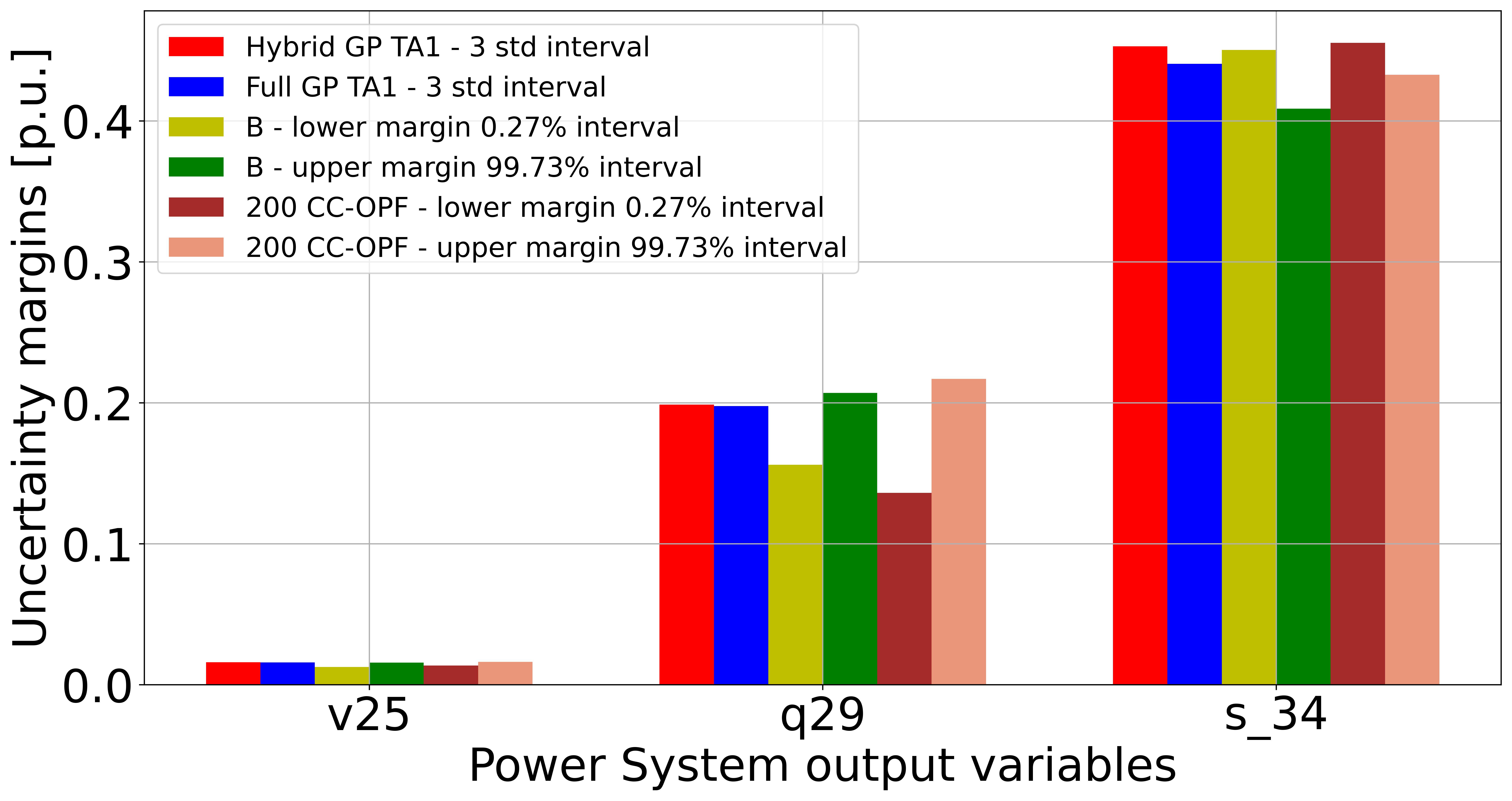}
 \caption{IEEE 9 case (left) and IEEE 39 case (right). Output uncertainty margins $\lambda$, for example, voltage, reactive and apparent power constraints.}%
 \label{fig:3}%
\end{figure}

Figure~\ref{fig:3} compares uncertainty margins \cite{schmidli2016stochastic} calculated with the hybrid and full GP CC-OPF (red and blue), the Monte Carlo simulation of AC-OPF for base-case (yellow and green), and the Monte Carlo simulation for CC-OPF solution (brown and salmon) with 100 (IEEE 9) and 200 (IEEE39) scenarios. The bar plots represent an example of voltage, reactive power, and apparent power flow uncertainty margins for each system. 

The 3 standard deviation distance for uncertainty margins is analytically derived using the estimated GP-variance (\ref{eq:27e}) around the estimated GP-mean (\ref{eq:27d}). The uncertainty margins from Monte Carlo are asymmetrical due to MC-based distributions from non-linear AC-OPF. Thus,  Monte Carlo uncertainty margins determine upper ($1-\epsilon$) = 99.73\% and lower ($\epsilon$) = 0.27\% quantiles (equal to 3 std) of the output distribution around the AC-PF solution $y(x)$, denoted by $y^{1-\epsilon}$ and $y^{\epsilon}$. Thus, the constraint tightenings are:
\begin{subequations}\label{eq:30}
\begin{align}
    \lambda^{upper} = y^{1-\epsilon} - y(x) \qquad  \lambda^{lower} = y(x) - y^{\epsilon}
\end{align}
\end{subequations}

The power margins are larger in absolute terms than the voltage margins. The GP CC-OPF margins of both approaches are symmetric, leading to larger or smaller margins compared with the upper and lower empirical quantiles obtained from the Monte Carlo AC-OPF and CC-OPF simulations in (\ref{eq:30}). This indicates that the GP uncertainty margins are well approximated. 

\subsection{Distributional Robustness}
In operational practice, a joint distribution of renewables and loads is not known exactly, and one needs to account for it when solving a problem. Furthermore, distribution mismatch might result from broken sensors or a cyberattack. This section shows that the proposed approach is robust against a sufficiently large mismatch between the expected distribution and the real one. Figure \ref{fig:1} illustrates the case of a corrupted distribution. Indeed, the interval $[7.5, 17]$ has no input data, leading to poor performance of simple GPR. However, the Hybrid GPR presented in this paper handles such intervals better. We also observe a similar result in the case of power grids.

Our experiment consists of three major steps: (1) change the data distribution adversarially, (2) train over the corrupted data, and (3) test the predictions of models. We compare the classical GPR \cite{mile} and the proposed hybrid GRR model. For an adversarial change of the input data distribution, we throw out a part of the input data within a certain interval; an example of this is shown in Fig.~III. 

We use IEEE 9 case with the dataset generated in Section \ref{sec:4} for our experiment. In each of the experiments, only one of the buses (bus 4, bus 6, and bus 8) is subjected to whipping out a 15 MW interval. See Fig.~III for details. Later on, we train the classical and the hybrid GPR on this data. Table~III illustrates a superior performance of the hybrid GPR on this data. 

\begin{figure*}[th]
 \centering
 \includegraphics[width=0.25\textwidth]{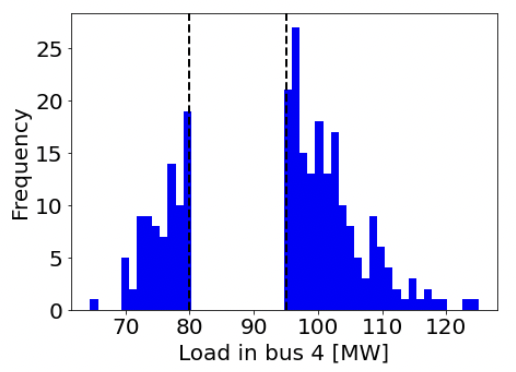} 
 \includegraphics[width=0.25\textwidth]{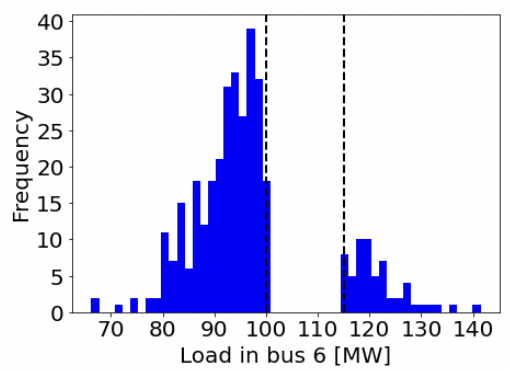}
 \includegraphics[width=0.25\textwidth]{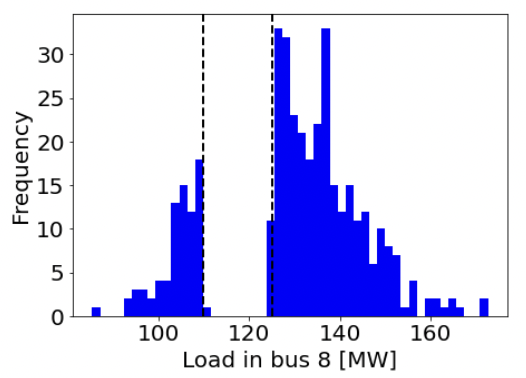}
    \caption{Histograms of buses' corrupted loads for each of the three experiments.}
\end{figure*}

\begin{table}[ht]
\centering
    \label{table:ieee9_RMSE_corrupt}
    \begin{tabular}{l|l|l|l}
        \hline\hline
        Experiment & RMSE GPR & RMSE Hybrid GPR & $\%$ dropped \\ \hline
        Load 4 & $3.65 \cdot 10^{-1}$ & $\boldsymbol {9.11 \cdot 10^{-3}}$ & $55.0\%$  \\ \hline
        Load 6 & $3.89 \cdot 10^{-1}$ & $\boldsymbol {11.60 \cdot 10^{-3}}$ & $38.0\%$ \\ \hline
        Load 8 & $9.25 \cdot 10^{-2}$ & $\boldsymbol {8.34 \cdot 10^{-3}}$ & $33.5\%$ \\ \hline
        \end{tabular}
    \caption{RMSE comparison for GPR and hybrid GPR over corrupted distributions. The last column specifies the portion of data distribution of the corresponding load that has been dropped.}
\end{table}


\section{Discussion} \label{sec:6}
The paper studies various machine learning approaches to chance-constrained AC optimal power flow problems. Because of the high non-linearity of the problem, we focused mostly on Gaussian Process Regression, one of the most popular model-free approaches to approximate non-linear stochastic data. Furthermore, it allows for avoiding spurious local minimums thus improving tractability of the AC-OPF problem. As the Gaussian Process Regression has poor scalability, we utilize a sparsity trick that allowed us to significantly reduce the complexity. Finally, we proposed a hybrid GRP approach that demonstrates higher robustness to missing and corrupted data than the classical GPR. 
\bibliographystyle{IEEEtran}
\bibliography{main.bib}





\vfill

\end{document}